\pdfoutput=1
\documentclass{article}
\usepackage{j}

\usepackage{tikz}
\usetikzlibrary{calc}
\usetikzlibrary{math}
\usetikzlibrary{decorations.pathreplacing}

\newlength{\barwidth}
\newcommand{\drawaxes}[1]{
  \draw[gray]  (-.2, 0) -> (3.75 * \barwidth + .2cm, 0);
  \draw                    (3.75 * \barwidth + .2cm, 0) node[anchor=west] {$x$};
  \draw[gray]  (-.2, 0) -> (-.2, 5.2);
  \draw                    (-.2, 5.2) node[anchor=south] {$#1$};
  \foreach \x in {0.5, 1.5, 2.5, 3.5}
    \draw[gray] (\x * \barwidth, -2pt) -- (\x * \barwidth, 0pt);
  \foreach \x/\char in {0.5/a, 1.5/b, 2.5/c, 3.5/d}
    \draw (\x * \barwidth, 0) node[anchor=north, text height=8] {\texttt\char};
}

\newcommand{\ytick}[2]{
  \draw[gray] (-.2cm - 2pt, #1) node[anchor=east, color=black] {#2} -- (-.2cm + 0pt, #1);
}

\newcommand{\drawbar}[3]{
  \draw[#3] (#1, 0) -- (#1, 12.5cm * #2);
}

\newcommand{\drawpmf}{
\begin{tikzpicture}
  \drawaxes{P(x)}
  \foreach \y in {0.125, 0.25, 0.375}{\ytick{12.5 * \y}{$\y$}}
  \drawbar{0.5\barwidth}{0.125}{}
  \drawbar{1.5\barwidth}{0.25}{}
  \drawbar{2.5\barwidth}{0.375}{}
  \drawbar{3.5\barwidth}{0.25}{}
\end{tikzpicture}
}

\newcommand{\drawapprox}[2]{
\begin{tikzpicture}
  \tikzmath{
    \splus = #1 + 1;
    \yunit = 12.5 / \splus;
  }
    \drawaxes{s'}
  \foreach \y [parse=true, count=\s] in {\yunit, 2 * \yunit, ..., 5.2}{
    \ytick{\y}{}
    \draw (-.2cm, \y - \yunit / 2) node[anchor=east]
      {\pgfmathparse{int(\s - 1)}#2$\pgfmathresult$};
  }
  \foreach \x/\freq/\start in {0/1/0, 1/2/1, 2/3/3, 3/2/6}{
    \foreach \y [parse=true, count=\s] in {\yunit, 2 * \yunit, ..., 5.2}{
      \tikzmath{
        \snew = int(8 * int((\s - 1) / \freq) + mod((\s - 1), \freq) + \start);
        if \snew<=#1 then {\toprint = \snew;} else {let \toprint = ;};
        }
      \draw ($(0, -\yunit / 2) + 0.5*(\barwidth,0)
      + \s*(0, \yunit) + \x*(\barwidth, 0)$) node {#2$\toprint$};
    }
  }
\end{tikzpicture}
}

\newlength{\intervalwidth}
\newlength{\gap}
\newlength{\lineheight}
\newcommand{\drawinterval}{
\begin{tikzpicture}
  \draw (-30pt, 0) node [anchor=east] {$s$};
  \draw (0, -.5\lineheight) -- (0, .5\lineheight);
  \draw (\intervalwidth, -.5\lineheight) -- (\intervalwidth, .5\lineheight);
  \foreach \s in {1, ..., 7}{
    \draw (\s * 0.125 * \intervalwidth, -.5\lineheight)
      -- (\s * 0.125 * \intervalwidth, .5\lineheight);
  }
  \foreach \y in {0, ..., 7}{
    \tikzmath{
      \x = (\y * 0.125 + 0.0625);
      \r = int(\y + 64);
    }
    \draw (\x\intervalwidth, 0) node {$\r$};
  }

  \draw (-30pt, -\lineheight -\gap) node [anchor=east] {$s \bmod 2^{r_p}$};
  \draw (0, -.5\lineheight - \gap) -- (0, -1.5\lineheight -\gap);
  \draw (\intervalwidth, -.5\lineheight - \gap)
    -- (\intervalwidth, -1.5\lineheight - \gap);
  \foreach \s in {1, ..., 7}{
    \draw (\s * 0.125 * \intervalwidth, -.5\lineheight - \gap)
      -- (\s * 0.125 * \intervalwidth, -1.5\lineheight - \gap);
  }
  \foreach \s in {0, ..., 7}{
    \tikzmath{
      \x = (\s * 0.125 + 0.0625);
    }
    \draw (\x\intervalwidth, -\lineheight - \gap) node  {$\s$};
  }

  \draw (-30pt, -2\gap - 2\lineheight) node [anchor=east] {$x$};
  \draw (0, -2\gap - 1.5\lineheight) -- (0, -2\gap - 2.5\lineheight - 6pt);
  \draw (\intervalwidth, -2\gap - 1.5\lineheight)
    -- (\intervalwidth, -2\gap - 2.5\lineheight - 6pt);
  \draw (0, -2\gap - 2.5\lineheight - 6pt) node [anchor=north] {$0$};
  \draw (\intervalwidth, -2\gap - 2.5\lineheight - 6pt) node [anchor=north] {$1$};
  \draw (0.125  \intervalwidth, -2\gap - 1.5\lineheight)
    -- (0.125 * \intervalwidth, -2\gap - 2.5\lineheight);
  \draw (0.375  \intervalwidth, -2\gap - 1.5\lineheight)
    -- (0.375 * \intervalwidth, -2\gap - 2.5\lineheight);
  \draw (0.75  \intervalwidth, -2\gap - 1.5\lineheight)
    -- (0.75 * \intervalwidth, -2\gap - 2.5\lineheight);
  \draw (0.06125\intervalwidth, -2\gap - 2\lineheight)
    node [text height=1ex] {\texttt{a}};
  \draw (0.25   \intervalwidth, -2\gap - 2\lineheight)
    node [text height=1ex] {\texttt{b}};
  \draw (0.5625 \intervalwidth, -2\gap - 2\lineheight)
    node [text height=1ex] {\texttt{c}};
  \draw (0.875  \intervalwidth, -2\gap - 2\lineheight)
    node [text height=1ex] {\texttt{d}};

\end{tikzpicture}
}

\newlength{\tbase}
\newcommand{\drawmessage}{
\begin{tikzpicture}
  \node at (0, 0) (s) {\texttt{0110001001110010}};
  \node at (0, -0.85) {\large\(s\)};
  \draw [{|-|}] (-1.5, 0.5) -- (1.5, 0.5);
  \node at (0, 0.8) {\(r_s\)};

  \node at (4.5, \tbase) {\texttt{01100010}};
  \node at (4.5, \tbase + 11) {\texttt{10100011}};
  \node at (4.5, \tbase + 30) {\(\vdots\)};
  \node at (4.5, \tbase + 43) {\texttt{11011001}};
  \draw [{|-|}] (4.5 - 0.75, \tbase + 58) -- (4.5 + 0.75, \tbase + 58);
  \node at (4.5, \tbase + 66) {\(r_t\)};
  \node at (4.5, -0.85) {\large\(t\)};
  \draw [{|-|}] (4.5 + 1.2, \tbase - 4) -- (4.5 + 1.2, \tbase + 43 + 4);
  \node at (4.5 + 1.6, \tbase + 21.5) {\(\abs{t}\)};
\end{tikzpicture}
}

\usepackage{listings}
\theoremstyle{definition}
\newtheorem{definition}{Definition}

\newcommand{\push}{\texttt{push}}
\newcommand{\pop}{\texttt{pop}}

\author{
  James Townsend\\
  University College London\\
  \texttt{james.townsend@cs.ucl.ac.uk}
}
\title{A tutorial on the range variant of asymmetric numeral systems}

\begin{document}
\maketitle
\begin{abstract}
  This paper is intended to be a brief and accessible introduction to the range
  variant of asymmetric numeral systems (ANS), a system for lossless
  compression of sequences which can be used as a drop in replacement for
  arithmetic coding (AC). Because of the relative simplicity of ANS, we are
  able to provide enough mathematical detail to rigorously prove that ANS
  attains a compression rate close to the Shannon limit. Pseudo-code, intuitive
  interpretation and diagrams are given alongside the mathematical derivations.
  A working Python demo which accompanies this tutorial is available at
  \url{https://raw.githubusercontent.com/j-towns/ans-notes/master/rans.py}.
\end{abstract}

\section{Introduction}\label{sec:intro}
  We are interested in algorithms for lossless compression of sequences of
  data.  Arithmetic coding (AC) and the range variant of asymmetric numeral
  systems (sometimes abbreviated to rANS, we simply use ANS) are examples of
  such algorithms. Just like arithmetic coding, ANS is close to optimal in
  terms of compression rate \citep{witten1987, duda2009}. The key difference
  between ANS and AC is in the order in which data are \emph{decoded}: in ANS,
  compression is last-in-first-out (LIFO), or `stack-like', while in AC it is
  first-in-first-out (FIFO), or `queue-like'. We recommend
  \citet[Chapter~4-6]{mackay2003} for background on source coding and arithmetic
  coding in particular. In this paper we will focus solely on ANS, which is not
  covered in detail by existing textbooks (although \cite{mcanlis2016} does
  briefly cover ANS).

  ANS comprises two basic functions, which we denote \push\ and \pop, for
  encoding and decoding, respectively (the names refer to the analogous stack
  operations). The \push\ function accepts some pre-compressed information
  \(m\) (short for `message'), and a symbol \(x\) to be compressed, and returns
  a new compressed message, \(m'\). Thus it has the signature
  \begin{equation}
    \push:(m, x) \mapsto m'.
  \end{equation}
  The new compressed message, \(m'\), contains precisely the same information
  as the pair \((m, x)\), and therefore \push\ can be inverted to form a
  decoder mapping.  The decoder, \pop, maps from \(m'\) back to \(m, x\):
  \begin{equation}
    \pop:m' \mapsto (m, x).
  \end{equation}
  Because the functions \push\ and \pop\ are inverse to one another, we have
  \(\push(\pop(m))=m\) and \(\pop(\push(m, x)) = (m, x)\).

\subsection{Specifying the problem which ANS solves}\label{sec:prob-spec}
  In this section we first define some notation, then describe the problem
  which ANS solves in more detail and sketch the high level approach to solving
  it. In the following we use `\(\log\)' as shorthand for the base 2 logarithm,
  usually denoted `\(\log_2\)'.

  The functions \push\ and \pop\ will both require access to the probability
  distribution from which symbols are drawn (or an approximation thereof). To
  describe distributions we use notation similar to \citet{mackay2003}:
  \begin{definition}\label{def:ensemble}
    An \emph{ensemble} \(X\) with precision \(r\) is a triple \((x,
    \mathcal{A}_X, \mathcal{P}_X)\) where the \emph{outcome} \(x\) is the value
    of a random variable, taking on one of a set of possible values
    \(\mathcal{A}_X = \{a_1, \ldots, a_I\}\), and \(\mathcal{P}_X = \{p_1,
    \ldots, p_I\}\) are the \emph{integer-valued} probability weights with each
    \(p_i\in\{1, \ldots, 2^r\}\), each \(P(x=a_i) = p_i / 2^r\) and therefore
    \(\sum_{i=1}^Ip_i = 2^r\).
  \end{definition}
  Note that this definition differs from the definition in \citet{mackay2003}
  in that the probabilities are assumed to be \emph{quantized}  to some
  precision \(r\) (i.e.\ representable by fractions \(p_i/2^r\)), and we assume
  that none of the \(a_i\) have zero probability. Having probabilities in this
  form is necessary for the arithmetic operations involved in ANS (as well as
  AC). Note that if we use a high enough \(r\) then we can specify
  probabilities with a precision similar to that of typical floating point ---
  32-bit floating points for example contain 23 `fraction' bits, and thus would
  have roughly the same precision as our representation with \(r=23\).

  The `information content' of an outcome can be measured using the following:
  \begin{definition}
    The \emph{Shannon information content} of an outcome \(x\) is
    \begin{equation}
      h(x) := \log \frac{1}{P(x)}
    \end{equation}
  \end{definition}
  Given a sequence of ensembles \(X_1, \ldots, X_N\), we seek an algorithm
  which can encode any outcome \(x_1, \ldots, x_N\) in a binary message whose
  length is close to \(h(x_1, \ldots, x_N) = \log 1/P(x_1, \ldots, x_N)\).
  According to Shannon's source coding theorem it is not possible to losslessly
  encode data in a message with expected length less than \(\mathbb{E}[h(x)]\),
  thus we are looking for an encoding which is close to optimal in expectation
  \citep{shannon1948}. Note that the joint information content of the sequence
  can be decomposed:
  \begin{align}
    \label{eq:info-decomp1}
    h(x_1, \ldots, x_N)
      &= \log\frac{1}{P(x_1, \ldots, x_N)}\\
      \label{eq:info-decomp2}
      &= \sum_n \log\frac{1}{P(x_n \given x_1, \ldots, x_{n-1})}\\
      \label{eq:info-decomp3}
      &= \sum_n h(x_n \given x_1, \ldots, x_{n-1}).
  \end{align}

  Because it simplifies the presentation significantly, we focus first on the
  ANS \emph{decoder}, the reverse mapping which maps from a compressed binary
  message to the sequence \(x_1, \ldots, x_N\). This will be formed of a
  sequence of \(N\) \pop\ operations; starting with a message \(m_0\) we define
  \begin{align}
    m_n, x_n = \pop(m_{n-1})\qquad\text{for }n=1, \ldots, N.
  \end{align}
  where each \pop\ uses the conditional distribution \(X_n\given X_1, \ldots,
  X_{n-1}\). We will show that the message resulting from each \pop, \(m_n\),
  is effectively shorter than \(m_{n-1}\) by no more than \(h(x_n \given x_1,
  \ldots, x_{n-1}) + \epsilon\) bits, where \(\epsilon\) is a small constant
  which we specify below, and therefore the difference in length between
  \(m_0\) and \(m_N\) is no more than \(h(x_1, \ldots, x_N) + N\epsilon\), by
  \cref{eq:info-decomp1,eq:info-decomp2,eq:info-decomp3}.

  We will also show that \pop\ is a bijection whose inverse, \push, is
  straightforward to compute, and therefore an encoding procedure can easily be
  defined by starting with a very short base message and adding data
  sequentially using \push. Our guarantee about the effect of \pop\ on message
  length translates directly to a guarantee about the effect of \push, in that
  the increase in message length due to the sequence of \push\ operations is
  less than \(h(x_1, \ldots, x_N) + N\epsilon\).

\section{Asymmetric numeral systems}
  Having set out the problem which ANS solves and given a high level overview
  of the solution in \Cref{sec:intro}, we now go into more detail, firstly
  discussing the data structure we use for \(m\), then the \pop\ function and
  finally the computation of its inverse, \push.

\subsection{The structure of the message}\label{sec:message}
\begin{figure}[ht]
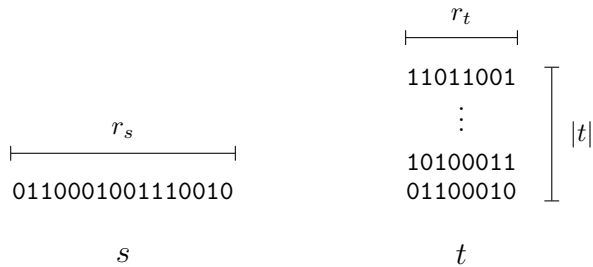

    \centering
    \drawmessage
    \caption{
      The two components of a message: the unsigned integer \(s\) (with \(r_s =
      16\)) and the stack of unsigned integers \(t\) (with \(r_t = 8\)). The
      integers are represented here in base 2 (binary).
  }
    \label{fig:message}
  \end{figure}

  We use a pair \(m = (s, t)\) as the data structure for the message \(m\). The
  element \(s\) is an unsigned integer with precision \(r_s\) (i.e.\ \(s \in
  \{0, 1, \ldots, 2^{r_s} - 1\}\), so that \(s\) can be expressed as a binary
  number with \(r_s\) bits). The element \(t\) is a stack of unsigned
  integers of some fixed precision \(r_t\) where \(r_t < r_s\).  This stack has
  its own push and pop operations, which we denote \(\texttt{stack\_push}\) and
  \(\texttt{stack\_pop}\) respectively. See \cref{fig:message} for a diagram of
  \(s\) and \(t\).  We need \(s\) to be large enough to ensure that our
  decoding is accurate, and so we also impose the constraint
  \begin{equation}\label{eq:s-constraint}
    s\geq2^{r_s - r_t},
  \end{equation}
  more detail on how and why we do this is given below. In the demo
  implementation we use \(r_s = 64\) and \(r_t = 32\).

  Note that a message can be flattened into a string of bits by concatenating
  \(s\) and the elements of \(t\). The length of this string is
  \begin{equation}
    l(m) := r_s + r_t\abs{t}
  \end{equation}
  where \(\abs{t}\) is the number of elements in the stack \(t\). We refer to
  this quantity as the `length' of \(m\). We also define the useful quantity
  \begin{equation}
    l^*(m) := \log s + r_t\abs{t}
  \end{equation}
  which we refer to as the `effective length' of \(m\). Note that the
  constraint in \cref{eq:s-constraint} and the fact that \(s < 2^{r_s}\) imply
  that
  \begin{equation}\label{eq:effective-length}
    l(m) - r_t \leq l^*(m) < l(m)
  \end{equation}
  Intuitively \(l^*\) can be thought of as a precise measure of the size of
  \(m\), whereas \(l\), which is integer valued, is a more crude measure.
  Clearly \(l\) is ultimately the measure that we care most about, since it
  tells us the size of a binary encoding of \(m\), and we use \(l^*\) to prove
  bounds on \(l\).

\subsection{Constructing the pop operation}
  To avoid notational clutter, we begin by describing the \pop\ operation for a
  single ensemble \(X = (x, \mathcal{A}_X, \mathcal{P}_X)\) with precision
  \(r\), before applying \pop\ to a sequence in \Cref{seq:pop-seq}. Our
  strategy for performing a decode with \pop\ will be firstly to extract a
  symbol from \(s\). We do this using a bijective function \(d:\mathbb
  N\rightarrow\mathbb N\times\mathcal{A}\), which takes an integer \(s\) as
  input and returns a pair \((s', x)\), where \(s'\) is an integer and \(x\) is
  a symbol. Thus \pop\ begins
  \begin{lstlisting}
def pop($m$):
    $s$, $t$ := $m$
    $s'$, $x$ := $d(s)$
  \end{lstlisting}
  We design the function \(d\) so that if \(s\geq 2^{r_s - r_t}\), then
  \begin{equation}\label{eq:inner-decoder}
    \log s - \log s' \leq h(x) + \epsilon
  \end{equation}
  where
  \begin{equation}
    \epsilon := \log \frac{1}{1 - 2^{-(r_s - r_t - r)}}.
  \end{equation}
  We give details of \(d\) and prove \cref{eq:inner-decoder} below. Note that
  when the term \(2^{-(r_s - r_t - r)}\) is small, the following approximation
  is accurate:
  \begin{equation}
    \epsilon \approx \frac{2^{-(r_s - r_t - r)}}{\ln 2}
  \end{equation}
  and thus $\epsilon$ itself is small. We typically use \(r_s = 64\), \(r_t =
  32\), and \(r = 16\), which gives \(\epsilon = \log 1/(1 - 2^{-16}) \approx
  2.2 \times 10^{-5}\).

  After extracting a symbol using \(d\), we check whether \(s'\) is below
  \(2^{r_s - r_t}\), and if it is we \texttt{stack\_pop} integers from \(t\)
  and move their contents into the lower order bits of \(s'\). We refer to this
  as `renormalization'.  Having done this, we return the new message and the
  symbol \(x\). The full definition of \pop\ is thus
  \begin{lstlisting}[frame=single]
def pop($m$):
    $s$, $t$ := $m$
    $s'$, $x$ := $d$($s$)
    $s$, $t$ := renorm($s'$, $t$) $\quad\leftarrow\textrm{ this function is defined below}$
    return ($s$, $t$), $x$
  \end{lstlisting}

  Renormalization is necessary to ensure that the value of \(s\) returned by
  \pop\ satisfies \(s\geq2^{r_s - r_t}\) and is therefore large enough that
  \cref{eq:inner-decoder} holds at the start of any future \pop\ operation. The
  \texttt{renorm} function has a while loop, which pushes elements from \(t\)
  into the lower order bits of \(s\) until \(s\) is full to capacity. To be
  precise:
  \begin{lstlisting}[frame=single]
def renorm($s$, $t$):
    # while $s$ has space for another element from $t$
    while $s < 2^{r_s - r_t}$:
        # pop an element $t_\mathrm{top}$ from $t$
        $t$, $t_{\mathrm{top}}$ := stack_pop($t$)

        # and push $t_\mathrm{top}$ into the lower bits of $s$
        $s$ := $2^{r_t} \cdot s$ + $t_{\mathrm{top}}$
    return $s$, $t$
  \end{lstlisting}

  The condition \(s < 2^{r_s - r_t}\) guarantees that \(2^{r_t} \cdot s +
  t_{\text{top}} < 2^{r_s}\), and thus there can be no loss of information
  resulting from overflow. We also have
  \begin{equation}
    \log (2^{r_t} \cdot s + t_\text{top}) \geq r_t + \log s
  \end{equation}
  since \(t_{\text{top}} \geq 0\).  Applying this inequality repeatedly, once
  for each iteration of the while loop in \texttt{renorm}, we have
  \begin{equation}\label{eq:renorm}
    \log s \geq \log s' + r_t\cdot\left[\text{\# elements popped from
    \(t\)}\right]
  \end{equation}
  where \(s, t = \texttt{renorm}(s', t)\) as in the definition of \pop.

  Combining \cref{eq:inner-decoder} and \cref{eq:renorm} gives us
  \begin{equation}\label{eq:pop-inequality}
    l^*(m) - l^*(m') \leq h(x) + \epsilon
  \end{equation}
  where \((m', x) = \pop(m)\), using the definition of \(l^*\). That is, the
  reduction in the effective message length resulting from \pop\ is close to
  \(h(x)\).

\subsection{Popping in sequence}\label{seq:pop-seq}
  We now apply \pop\ to the setup described in \Cref{sec:prob-spec}, performing
  a sequence of \pop\ operations to decode a sequence of data. We suppose that
  we are given some initial message \(m_0\).

  For \(n=1\ldots N\), we let \(m_n, x_n = \pop(m_{n-1})\) as in
  \Cref{sec:prob-spec}, where each \pop\ uses the corresponding distribution
  \(X_n \given X_1, \ldots, X_{n-1}\). Applying \cref{eq:pop-inequality} to
  each of the \(N\) \pop\ operations, we have:
  \begin{align}
    l^*(m_0) - l^*(m_N)
      &= \sum_{n=1}^N [l^*(m_{n-1}) - l^*(m_n)]\\
      &\leq \sum_{n=1}^N [h(x_n \given x_1, \ldots, x_{n-1}) + \epsilon]\\
      &\leq h(x_1, \ldots, x_N) + N\epsilon\label{eq:effective-lengths}
  \end{align}

  This result tells us about the reduction in message length from \pop\, but
  also, conversely, about the length of a message \emph{constructed} using
  \push. We can actually initialize an encoding procedure by \emph{choosing}
  \(m_N\), and then performing a sequence of \push\ operations. Since our
  ultimate goal when encoding is to minimize the encoded message length \(m_0\)
  we choose the setting of \(m_N\) which minimizes \(l^*(m_N)\), which is \(m_N
  = (s_N, t_N)\) where \(s_N = 2^{r_s - r_t}\) and \(t_N\) is an empty stack.
  That gives \(l^*(m_N) = r_s - r_t\) and therefore, by
  \cref{eq:effective-lengths},
  \begin{equation}
    l^*(m_0) \leq h(x_1, \ldots, x_N) + N\epsilon + r_s - r_t.
  \end{equation}
  Combining that with \cref{eq:effective-length} gives an expression for the
  actual length of the flattened binary message resulting from \(m_0\):
  \begin{equation}
    l(m_0) \leq h(x_1, \ldots, x_N) + N\epsilon + r_s.
  \end{equation}

  It now remains for us to describe the function \(d\) and show that it
  satisfies \cref{eq:inner-decoder}, as well as showing how to invert \pop\ to
  form the encoding function \push.

\subsection{The function \(d\)}
The function \(d:\mathbb{N}\rightarrow\mathbb{N}\times\mathcal{A}\) must be a
bijection, and we aim for \(d\) to satisfy \cref{eq:inner-decoder}, and thus
\(P(x)\approx \frac{s'}{s}\).  Achieving this is actually fairly
straightforward.  One way to define a bijection \(d:s\mapsto(s', x)\) is to
start with a mapping \(\tilde d: s\mapsto x\), with the property that none of
the preimages \(\tilde d^{-1}(x):=\{n\in\mathbb{N}:\tilde d(n) = x\}\) are
finite for \(x\in\mathcal{A}\). Then let \(s'\) be the index of \(s\) within
the (ordered) set \(\tilde d^{-1}(x)\), with indices starting at \(0\).
Equivalently, \(s'\) is the number of integers \(n\) with \(0\leq n<s\) and
\(d(n) = x\).

With this setup, the ratio
\begin{equation}\label{eq:ratio}
  \frac{s'}{s} = \frac{\abs{\{n\in\mathbb{N}: n < s, d(n) = x\}}}{s}
\end{equation}
is the density of numbers which decode to \(x\), within all the natural numbers
less \(s\). For large \(s\) we can ensure that this ratio is close to \(P(x)\)
by setting \(\tilde d\) such that numbers which decode to a symbol \(x\) are
distributed \emph{within the natural numbers} with density close to \(P(x)\).

To do this, we partition \(\mathbb{N}\) into finite ranges of equal length, and
treat each range as a model for the interval \([0, 1]\), with sub-intervals
within \([0, 1]\) corresponding to each symbol, and the width of each
sub-interval being equal to the corresponding symbol's probability (see
\cref{fig:interval}). To be precise, the mapping \(\tilde d\) can then be
expressed as a composition \(\tilde d = \tilde d_2 \circ \tilde d_1\), where
\(\tilde d_1\) does the partitioning described above, and \(\tilde d_2\)
assigns numbers within each partition to symbols (sub-intervals). So
\begin{equation}
  \tilde d_1(s) := s \bmod 2^{r}.
\end{equation}
Using the shorthand \(\bar{s} := \tilde d_1 (s)\), and defining
\begin{equation}
  c_j := \begin{cases}
    0                    &\quad\text{if }j=1\\
    \sum_{k=1}^{j-1} p_k &\quad\text{if }j=2,\ldots,I
  \end{cases}
\end{equation}
as the (quantized) cumulative probability of symbol \(a_{j-1}\),
\begin{equation}
  \tilde d_2(\bar s) := a_i\text{ where }i := \max \{j : c_j \leq \bar s\}.
\end{equation}
That is, \(\tilde d_2(\bar s)\) selects the symbol whose sub-interval contains
\(\bar s\).  \Cref{fig:interval} illustrates this mapping, with a particular
probability distribution, for the range \(s = 64,\ldots, 71\).

\begin{figure}[ht]
  \centering
  \drawinterval
  \caption{
    Showing the correspondence between \(s\), \(s \bmod 2^{r}\) and the
    symbol \(x\). The interval \([0, 1]\subset\mathbb{R}\) is modelled by the
    set of integers \(\{0, 1, \ldots, 2^{r} - 1\}\). In this case \(r = 3\)
    and the probabilities of each symbol are \(P(\mathtt a) =
    \nicefrac{1}{8}\), \(P(\mathtt b) = \nicefrac{2}{8}\), \(P(\mathtt c) =
    \nicefrac{3}{8}\) and \(P(\mathtt d) = \nicefrac{2}{8}\).}
  \label{fig:interval}
\end{figure}

\begin{figure}[ht]
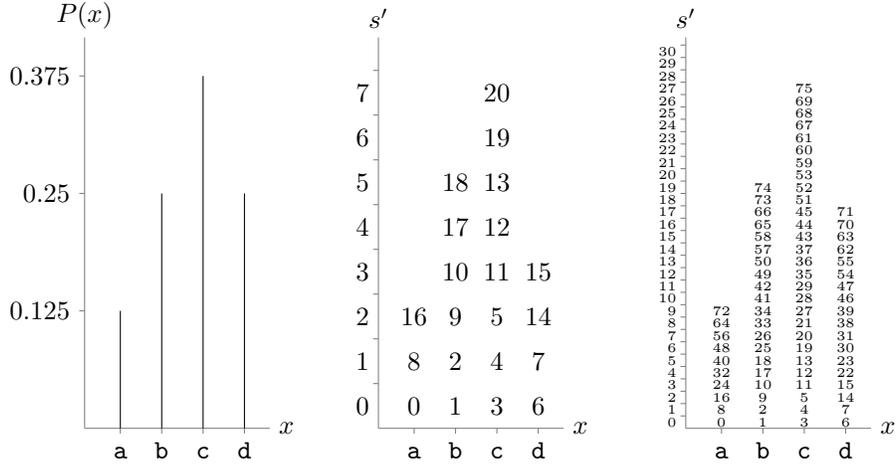

  \centering
  \drawpmf \quad \drawapprox{20}{} \quad \drawapprox{75}{\tiny}
  \caption{
    Showing the pmf of a distribution over symbols (left) and a visualization
    of the mapping \(d\) (middle and right). In the middle and right figures,
    numbers less than or equal to \(s_\mathrm{max}\) are plotted, for
    \(s_\mathrm{max}=20\) and \(s_\mathrm{max}=75\).  The position of each
    number \(s\) is set to the coordinates \((x, s')\), where \(s', x = d(s)\).
    The heights of the bars are thus determined by the ratio \(s'/s\) from
    \cref{eq:ratio}, and can be seen to approach the heights of the lines in
    the histogram on the left (that is, to approach \(P(x)\)) as the density of
    numbers increases.
  }\label{fig:visual-ans}
\end{figure}

\subsection{Computing \(s'\)}
The number \(s'\) was defined above as ``the index of \(s\) within the
(ordered) set \(\tilde d^{-1}(x)\), with indices starting at \(0\)''. We now
derive an expression for \(s'\) in terms of \(s\), \(p_i\) and \(c_i\), where
\(i = \max\{j: c_j \leq \bar s\}\) (as above), and we prove
\cref{eq:inner-decoder}.

Our expression for \(s'\) is a sum of two terms. The first term counts the
entire intervals, corresponding to the selected symbol \(a_i\), which are below
\(s\). The size of each interval is \(p_i\) and the number of intervals is
\(s\div 2^{r}\), thus the first term is \(p_i \cdot (s \div 2^{r})\), where
\(\div\) denotes \emph{integer} division, discarding any remainder.  The second
term counts our position within the current interval, which is \(\bar s - c_i
\equiv s\bmod 2^{r} - c_i\). Thus
\begin{equation}\label{eq:s' def}
  s' = p_i \cdot (s \div 2^{r}) + s\bmod 2^{r} - c_i.
\end{equation}
This expression is straightforward to compute. Moreover from this expression it
is straightforward to prove \cref{eq:inner-decoder}. Firstly, taking the
\(\log\) of both sides of \cref{eq:s' def} and using the fact that \(s\bmod
2^{r} - c_i \geq 0\) gives
\begin{align}
  \log s' \geq \log (p_i\cdot (s\div 2^{r})).
\end{align}
then by the definition of \(\div\), we have \(s\div 2^{r} > \frac{s}{2^{r}}
- 1\), and thus
\begin{align}
  \log s'
    &\geq \log\left(p_i\left(\frac{s}{2^{r}} -1\right)\right)\\
    &\geq \log s - h(x) + \log\left(1 - \frac{2^{r}}{s}\right)\\
    &\geq \log s - h(x) - \epsilon
\end{align}
as required, using the fact that \(P(x) = \frac{p_i}{2^{r}}\) and \(s \geq
2^{r_s - r_t}\).

By choosing \(r_s - r_t\) to be reasonably
large (it is equal to 32 in our implementation), we ensure that
\(\frac{s'}{s}\) is very close to \(P(x)\). This behaviour can be seen visually
in \cref{fig:visual-ans}, which shows the improvement in the approximation for
larger \(s\).

\subsection{Pseudocode for \(d\)}
We now have everything we need to write down a procedure to compute \(d\). We
assume access to a function \(f_X:\bar{s}\mapsto (a_i, c_i, p_i)\), where \(i\)
is defined above. This function clearly depends on the distribution of \(X\), and
its computational complexity is equivalent to that of computing the CDF and
inverse CDF for \(X\). For many common distributions, the CDF and inverse CDF
have straightforward closed form expressions, which don't require an explicit
sum over \(i\).

\pagebreak
We compute \(d\) as follows:
\begin{lstlisting}[frame=single]
def $d$($s$):
    $\bar s$ := $s\bmod 2^{r}$
    $x$, $c$, $p$ := $f_X(\bar s)$
    $s'$ := $p \cdot (s \div 2^{r}) + \bar s - c$
    return $s'$, $x$
\end{lstlisting}

\subsection{Inverting the decoder}
Having described a decoding process which appears not to throw away any
information, we now derive the inverse process, \push, and show that it is
computationally straightforward.

The \push\ function has access to the symbol \(x\) as one of its inputs, and
must do two things. Firstly it must \texttt{stack\_push} the correct number of
elements to \(t\) from the lower bits of \(s\). Then it must reverse the effect
of \(d\) on \(s\), returning a value of \(s\) identical to that before \pop\
was applied.

Thus, on a high level, the inverse of the function \pop\ can be expressed as
\begin{lstlisting}[frame=single]
def push($m$, $x$):
    $s$, $t$ := $m$
    $p$, $c$ := $g_X$($x$)
    $s'$, $t$ := renorm_inverse($s$, $t$; $p$)
    $s$ := $d^{-1}$($s'$; $p$, $c$)
    return $s$, $t$
\end{lstlisting}
where \(g_X:x\mapsto (p_i, c_i)\) with \(i\) as above. The function \(g_X\) is
similar to \(f_X\) in that it is analogous to computing the quantized CDF and
mass function \(x \mapsto p_i\).  The function \(d^{-1}\) is really a
pseudo-inverse of \(d\); it is the inverse of \(s\mapsto d(s, x)\), holding
\(x\) fixed.

As mentioned above, \texttt{renorm\_inverse} must \texttt{stack\_push} the
correct amount of data from the lower order bits of \(s\) into \(t\). A
necessary condition which the output of \texttt{renorm\_inverse} must satisfy
is
\begin{equation}\label{eq:renorm_inverse_ineq}
  2^{r_s - r_t} \leq d^{-1}(s'; p, c) < 2^{r_s}
\end{equation}
This is because the output of \push\ must be a valid message, as described in
\Cref{sec:message}, just as the output of \pop\ must be.

The expression for \(s'\) in \cref{eq:s' def} is straightforward to invert,
yielding a formula for
\(d^{-1}\):
\begin{equation}
  d^{-1}(s'; p, c) = 2^{r} \cdot (s' \div p) + s' \bmod p + c
\end{equation}
We can substitute this into \cref{eq:renorm_inverse_ineq} and simplify:
\begin{align}
      &&2^{r_s - r_t}            &\leq 2^{r} \cdot (s' \div p) + s' \bmod p + c < 2^{r_s}\\
  \iff&&2^{r_s - r_t}            &\leq 2^{r} \cdot (s' \div p) < 2^{r_s}\\
  \iff&&p\cdot2^{r_s - r_t - r}&\leq s' < p\cdot 2^{r_s - r}\label{eq:simple-cond}
\end{align}
So \texttt{renorm\_inverse} should move data from the lower order bits of
\(s'\) into \(t\) (decreasing \(s'\)) until \cref{eq:simple-cond} is satisfied.
To be specific:

\begin{lstlisting}[frame=single]
def renorm_inverse($s'$, $t$; $p$):
    while $s' \geq p \cdot 2^{r_s - r}$:
        $t$ := stack_push($t$, $s'\bmod 2^{r_t}$)
        $s'$ := $s'\div 2^{r_t}$
    return $s'$, $t$
\end{lstlisting}

Although, as mentioned above, \cref{eq:simple-cond} is a \emph{necessary}
condition which \(s'\) must satisfy, it isn't immediately clear that it's
sufficient. Is it possible that we need to continue the while loop in
\texttt{renorm\_inverse} past the first time that \(s'<p\cdot2^{r_s - r}\)? In
fact this can't be the case, because \(s'\div2^{r_t}\) decreases \(s'\) by a
factor of at least \(2^{r_t}\), and thus as we iterate the loop above we will
land in the interval specified by \cref{eq:simple-cond} at most once. This
guarantees that the \(s\) that we recover from \texttt{renorm\_inverse} is the
correct one.

\section{Further reading}
Since its invention by \citet{duda2009}, ANS appears not to have gained
widespread attention in academic literature, despite being used in various
state of the art compression systems. At the time of writing, a search on
Google Scholar for the string ``asymmetric numeral systems" yields 148 results.
For comparison, a search for ``arithmetic coding", yields `about 44,000'
results. As far as I'm aware, ANS has appeared in only one textbook, with a
practical, rather than mathematical, presentation \citep{mcanlis2016}.

However, for those wanting to learn more there is a huge amount of material on
different variants of ANS in \citet{duda2009} and \citet{duda2015}. A
parallelized implementation based on SIMD instructions can be found in
\citet{giesen2014} and a version which encrypts the message whilst compressing
in \citet{duda2016}.  An extension of ANS to models with latent variables was
developed by \citet{townsend2019}.

Duda maintains a list of ANS implementations at
\url{
  https://encode.su/
  threads/2078-List-of-Asymmetric-Numeral-Systems-implementations}.

\subsection*{Acknowledgements}
  Thanks to Tom Bird for feedback on drafts of this paper. Thanks also to
  Szymon Grabowski for pointing out that ANS is covered in \citet{mcanlis2016}.

\printbibliography
\end{document}